\documentclass[epj]{pw}
\usepackage{graphics}
\usepackage{graphicx}
\usepackage{amsmath}
\usepackage{amssymb}
\usepackage{color}
\begin{document}
\title{Echo Spectroscopy of Atomic Dynamics in a Gaussian Trap via Phase Imprints}
\author{Daniel Oblak \and J\"{u}rgen Appel \and Patrick Windpassinger   \and Ulrich Busk Hoff \and Niels Kj{\ae}rgaard\thanks{\email{kjaergaard@nbi.dk}} \and Eugene S. Polzik
}                     
%
%
\institute{QUANTOP, Danish National Research Foundation Center for
Quantum Optics, Niels Bohr Institute, University of Copenhagen,
Blegdamsvej 17, DK-2100 Copenhagen \O, Denmark}
\date{Dated: \today}
%
\abstract{We report on the collapse and revival of Ramsey fringe
visibility when a spatially dependent phase is imprinted in the
coherences of a trapped ensemble of two-level atoms. The phase is
imprinted via the light shift from a Gaussian laser beam which
couples the dynamics of internal and external degrees of freedom for
the atoms in an echo spectroscopy sequence. The observed revivals
are directly linked to the oscillatory motion of atoms in the trap.
An understanding of the effect is important for quantum state
engineering of trapped atoms.
} 
\maketitle
\section{Introduction}\label{intro}
A trapped gas of atoms can act as dispersive medium
with a refractive index which depends on the internal state of the
atoms. The state-dependent phase shift of probe laser light
propagating through an ensemble of Cs atoms has recently been used
to observe Rabi flopping on the clock transition non-destructively
\cite{Chaudhury2006,Windpassinger2008}. Such non-destructive
measurements of a collective atomic quantum state component holds the
promise to predict the outcome of subsequent measurements
beyond the standard quantum limit \cite{Kuzmich1998}. This reduction
in uncertainty is referred to as conditional squeezing and the
resulting nonclassical atomic state may be used to increase the
precision of atomic clocks \cite{Oblak2005}

\begin{figure}[tb!]
\begin{center}
\includegraphics[width=0.8\columnwidth]{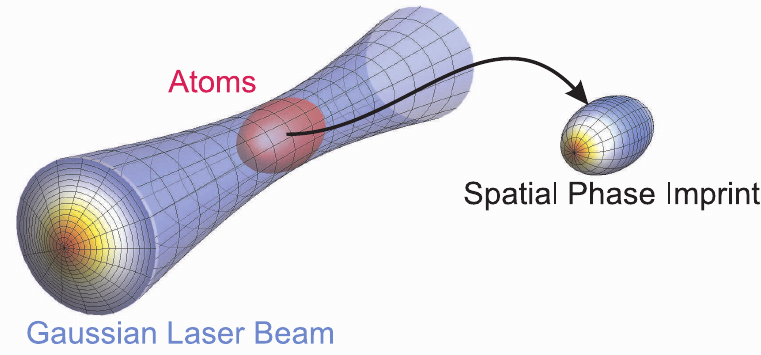}
\caption{The Gaussian intensity distribution of a laser beam is
mapped onto the atoms as a spatially dependent phase of the clock
state superpositions.}\label{fig:1}
\end{center}
\end{figure}

In a recent paper \cite{pwarxiv1}, we considered the effect of
inhomogeneous light shifts on the atomic quantum state evolution
when using a Gaussian laser beam for dispersive probing of an
ensemble of atoms confined in a dipole trap. The spatial intensity
distribution of the probe beam implies that an atom experiences a
position dependent \textit{differential} ac Stark shift of the clock
levels and the atomic cloud acquires a spatial phase imprint as
illustrated schematically in Fig.~\ref{fig:1}. In the present paper
we shall focus on the fact that the individual atoms are not
stationary in the trap, but move about to explore regions of
different probe light intensities. This is of consequence for how
well their inhomogeneous phase spread can be compensated for by
using Hahn echo techniques. Specifically, we shall investigate the
degradation of Ramsey fringe contrast in echo spectroscopy as a
result of atomic movement in between two perturbing light pulses on
either side of the echo pulse. The fringe contrast can be observed
to revive at half-integer multipla of the radial trap period and we
demonstrate how this effect can be used to measure the trap
frequency \textit{in situ} without actually exciting collective
oscillation modes. The specific form of the collapse and revival of
fringe visibility is modelled readily when the anharmonicity of the
trapping potential is taken into account.

\section{Ramsey Interrogation and Echo Spectroscopy}

\subsection{Ramsey spectroscopy of two-level atoms}
In Ramsey spectroscopy, a collection of two-level atoms with quantum
levels \mbox{$|\!\downarrow\rangle$} and \mbox{$|\!
\uparrow\rangle$} separated by an
energy difference $\hbar\Omega_0$ interacts with two near resonant
fields of angular frequency $\Omega\approx \Omega_0$ separated by a
time $\mathcal{T}$ \cite{Vanier1989}. With all the atoms initially
in the state \mbox{$|\!\downarrow\rangle$} the first field interaction --- a
$\pi/2$-pulse
--- produces an equal coherent superposition state
\mbox{$(|\!\downarrow\rangle+|\!\uparrow\rangle)/\sqrt{2}$} for each atom. The
atoms now evolve freely for the time $\mathcal{T}$ during which a
phase $\phi=(\Omega-\Omega_0)\mathcal{T}$ is accumulated by the state:
\mbox{$(|\!\downarrow\rangle\,+\, e^{i\phi}|\!\uparrow\rangle)/\sqrt{2}$}. Finally,
a second $\pi/2$-pulse is applied and the population difference of
the two states is measured. This quantity will vary periodically
with $\phi$ giving rise to so-called Ramsey fringes when either
$\mathcal{T}$ or $\Omega$ are scanned.

If the two energy levels \mbox{$|\!\downarrow\rangle$} and \mbox{$|\!\uparrow\rangle$}
are perturbed differentially in any way during the free evolution
period, the energy splitting $\hbar \Omega_0$ and thus $\phi$ will be affected.
This may occur homogeneously
such that all the atoms are perturbed by the same amount causing
an over all shift of the Ramsey fringe pattern \cite{Featonby1998}. Essentially, this
fringe shift constitutes an interferometric measurement of the
perturbation strength.
For nonuniform perturbations the phase $\phi$ differs from atom to atom and in the Ramsey measurement of the whole collection of atoms the fringe shift is accompanied by a degraded visibility as a result of \textit{inhomogeneous} dephasing \cite{pwarxiv1}.


\begin{figure}[t!]
\begin{center}
\includegraphics[width=\columnwidth]{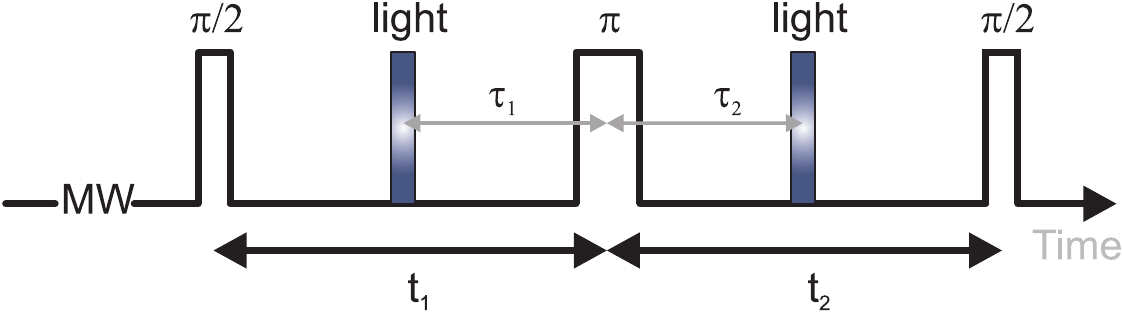} \caption{Echo
spectroscopy time line as employed in our experiments. A
near-resonant microwave (MW) field is applied to the atoms in a
$\pi/2-\pi-\pi/2$ pulse sequence. The effect of short perturbing
light pulses occurring at a time $t_1$ before and a time $t_2$ after
the $\pi$-pulse is investigated. }\label{fig:sequence}
\end{center}
\end{figure}

\subsection{Coherence echoes}
\begin{figure*}[tb!]
\begin{center}
\includegraphics[width=\textwidth]{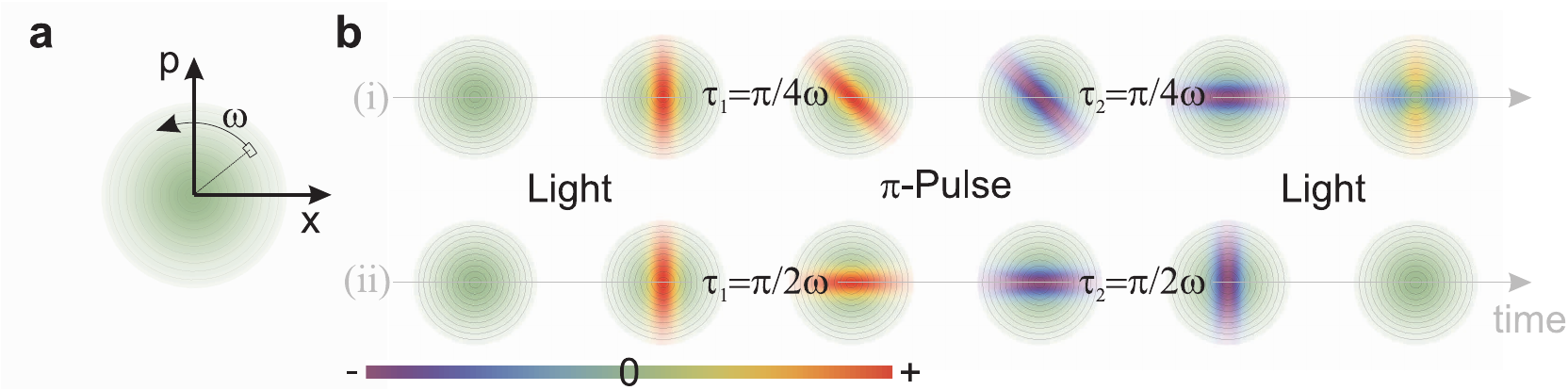}
\caption{(a) Dynamical phase space representation of atoms in a 1D
harmonic oscillator potential of angular frequency $\omega$. (b)
Phase imprint on the atoms from the position dependent light
intensity of Gaussian laser beam during echo spectroscopy, where two
identical light pulses act on each side of the echo $\pi$-pulse. (i)
In general echo spectroscopy cannot completely rephase the atoms.
(ii) In the special case of the time between the two light pulses
being $\tau_1+\tau_2=\pi({\rm mod}\pi)/\omega$, the effects of the
two light pulses cancel.}\label{fig:2}. \end{center}
\end{figure*}

Effects of any form of inhomogeneous dephasing encountered in Ramsey
spectroscopy can, to a large extent, be cancelled by introducing a
so-called echo-pulse ($\pi$-pulse) between the two $\pi/2$-pulses at
time $t_1$ in the Ramsey sequence \cite{HAHN1950}. The echo pulse essentially
inverts the sign of the phase $\phi$ accumulated so far by each atom and
if subsequently each atom encounters the same amount of perturbation as
prior to the $\pi$-pulse a rephasing  will occur at time $t_2=t_1$
after the echo pulse. In this sense the perturbation is \textit{reversible} e.g. for a collection of stationary atoms in a inhomogeneous light field.
Irreversible dephasing may result from fluctuating perturbations
such as the variations in phase shift due to noise in the intensity
of the inhomogeneous light field or movement of atoms therein
\cite{Kuhr2005}.

Since echo spectroscopy nullifies the effect of reversible perturbations it has proven to be a powerful tool for investigating the nature and magnitude of atomic decoherence due to the combined effect of irreversible processes including dephasing and spontaneous decay of atoms \cite{Kuhr2005,Andersen2003,Ozeri2005,Lengwenus2007}. Several studies focus on the influence of the inhomogeneous dipole trapping light and it has been demonstrated that an echo pulse is efficient in compensating the effect of the
trapping laser on a time scale $\ll h/\Delta E$, where $\Delta E$ is
the differential ac Stark shift between \mbox{$|\!\downarrow\rangle$} and
\mbox{$|\!\uparrow\rangle$} \cite{Andersen2003}. Furthermore, it has been
shown that the differential light shift for an oscillator mode of a
far-off-resonant trap (FORT) is well described by its time averaged
value \cite{Kuhr2005} such that inhomogeneous dephasing introduced
by the Gaussian trapping laser beam profile can in many cases be reversed in an
echo sequence.

\subsection{Inhomogeneous light shifts and trap
dynamics}\label{sec:theodynamics} In the present paper we shall
consider the perturbations from an auxiliary inhomogeneous pulse of
light applied before the echo pulse and an identical light pulse
applied after the echo pulse as indicated in
Fig.~\ref{fig:sequence}. For a stationary atomic ensemble perfect
rephasing would be expected except for decoherence of the atomic
state due to irreversible spontaneous scattering events and quantum
fluctuations (shot noise) of the two light pulses. Hence, a
measurement of the Ramsey fringe visibility would appear to be
ideally suited for determining the amount of decoherence introduced
by a given laser beam. The situation is somewhat complicated when
atoms change positions between the application of the two light
pulses. To elaborate on this issue we consider a simple 1D model of
particles evolving in a harmonic trap. The distributions of atoms in
momentum and poisition are represented as a dynamical phase space
plot [see Fig.~\ref{fig:2} (a)]. The application of a light pulse
with Gaussian intensity profile will imprint a phase which only
depends on the position if we assume that the duration of the light
pulse is much shorter than an oscillation period $T$ in the trap.
Now, during the free evolution the imprint will rotate in dynamical
phase space at angular frequency $\omega=2\pi/T$. In
Fig.~\ref{fig:2}(b), we outline the situation for echo spectroscopy
in the cases (i) $\tau_1=\tau_2=\pi/4\omega$ and (ii)
$\tau_1=\tau_2=\pi/2\omega$, where $\tau_1$ and $\tau_2$ are the
durations from the light pulse before and after to the echo pulse,
respectively. Complete rephasing of the atoms is encountered only if
$\tau_1+\tau_2=\pi({\rm mod}~\pi)/\omega$, i.e. the atoms oscillate
in the trap for a half-integer multipla of a trap period between the
two light pulses. For an echo spectroscopy experiment this implies
that the Ramsey fringe contrast will depend on the time separation
$\tau_1+\tau_2$ between the light pulses.

\section{Experimental}
\subsection{Setup}
Details of our experimental setup and atomic sample preparation can
found in \cite{pwarxiv1}. The starting point for the experiment
presented in the present paper is an ensemble of $\sim
10,000-50,000$ Cs atoms polarized in the
$6S_{1/2}(F=3,m_F=0)\equiv|\!\!\downarrow\rangle$ clock state and
confined by a $\sim 4~W$ Yb:YAG laser beam focussed to a waist of
$\rm \sim 40 \mu m$. This dipole trap is characterized by
oscillation frequencies in the order of a kHz radially and a few Hz
axially. The clock transition
\mbox{$|\!\downarrow\rangle$}~$\leftrightarrow6S_{1/2}(F=4,m_F=0)\equiv$~\mbox{$|\!\uparrow\rangle$}
is driven using microwave radiation around 9.2~GHz applied in a
$\pi/2-\pi-\pi/2$ echo spectroscopy sequence as depicted in
Fig.~\ref{fig:sequence}. At the end of this sequence we determine
the fraction of atoms residing in the $|\!\uparrow\rangle$ state. The
probing of atoms is performed with a beam of light propagating along
the trap axis with a waist of 18~$\rm \mu m$ located at the center
of the trap. The frequency of the probe light is blue detuned by
160~MHz from the {$6S_{1/2}(F=4)\rightarrow6P_{3/2}(F=5)$} transition
and via the dispersive atom-light interaction the probe light
experiences a phase shift proportional to the number of
\mbox{$|\!\uparrow\rangle$}-atoms which is measured using a shot noise
limited Mach Zehnder interferometer \cite{Oblak2005,Petrov2007a}. By
applying light resonant with the
{$6S_{1/2}(F=3)\rightarrow6P_{3/2}(F=4)$} transition all atoms are
pumped into the $6S_{1/2}(F=4)$ level and the total number of atoms
involved is determined from a subsequent phase shift measurement.
Optionally, we can apply light from an additional probe laser which
is red detuned by 135~MHz from the
{$6S_{1/2}(F=3)\rightarrow6P_{3/2}(F=2)$} transition and hence couples
to the \mbox{$|\!\downarrow\rangle$} population. By engaging the two probes
simultaneously we can obtain a zero (mean) interferometer phase
shift for ensembles in an equal clock state superposition
irrespective of the total number of atoms. This two probe color
configuration has proven convenient in our measurements of atomic
projection noise \cite{Windpassinger2008}. The two probes of this
two-color scheme are merged in a single mode optical fiber to ensure
good spatial overlap and hence enter the interferometer at the same
input port.

\subsection{Decoherence and inhomogeneous dephasing}
For low optical powers the dispersive probing scheme is close to
nondestructive in the sense that spontaneous scattering is very
limited. However, the ratio of signal to noise (which in our case is
the shot noise of light) for a phase shift measurements increases
with increasing probe photon number. Hence, there is a trade off
between information gained and coherence lost. Establishing an
optimal balance between decoherence and measurement strength is important for quantum state engineering of a squeezed
clock state population difference via a dispersive measurement. As it turns out, the
optical depth of the atomic sample is the key figure of merit
determining the optimal optical power and thus the amount of
decoherence \cite{Hammerer2004}. Preferably, and to achieve a high
degree of squeezing, the optical depth should be large such that
each probe photon interacts with many atoms. In this respect the
$\sim 1:200$ radial to axial aspect ratio of our sample provided by
the dipole trap potential is favorable and gives rise to an optical
depth of up to $\sim 20$ along the direction of the probe laser
beam.

From these considerations, it is obviously important to have a
handle on the amount of decoherence introduced by the dispersive
probing scheme and echo spectroscopy would appear to be the method
of choice \cite{Ozeri2005}. Ideally, a light pulse derived from the
probe laser could be devided into two with each part applied before
and after the echo pulse, respectively, and the reduction in Ramsey
fringe amplitude would then gauge the decoherence from probe light
considered as a perturbation. This method relies on the complete cancellation of the reversible dephasing by the inhomogeneous light shift. However, the movement of atoms in between the two
pertubing pulses may also lead to imperfect rephasing causing a Ramsey
fringe reduction as outlined in \ref{sec:theodynamics}. In fact,
this effect is significant and is prominently manifested in the echo spectroscopy.

\subsection{Results}
\subsubsection{Echo Spectroscopy}
\begin{figure}[b!]
\begin{center}
\includegraphics[width=\columnwidth]{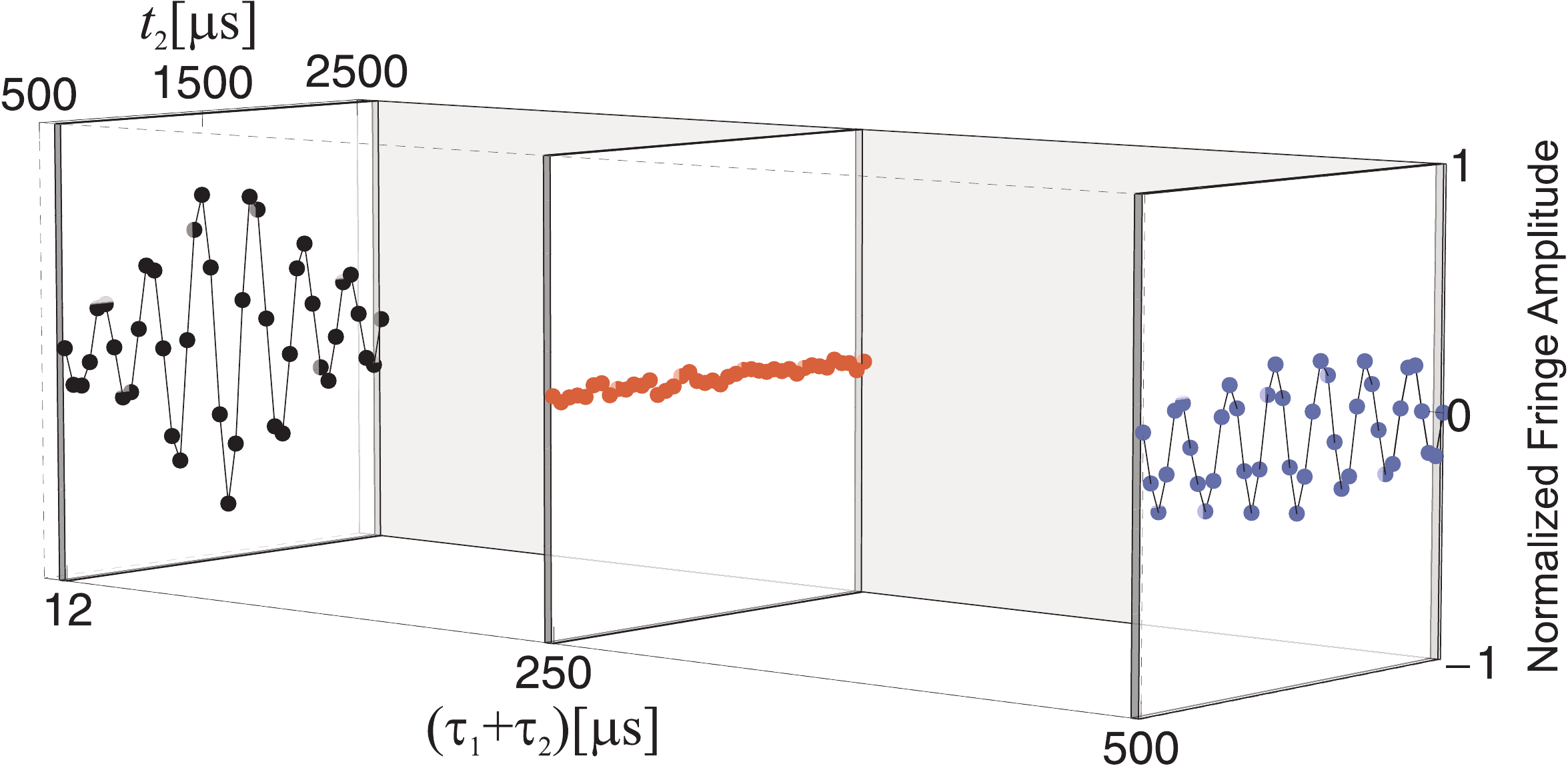}
\caption{Ramsey fringes as recorded in echo spectroscopy as a
function of the time separation $\tau_1+\tau_2$ between two light
shifting pulses around the echo pulse. The fringe amplitudes have
been normalized to the case when no light is applied. For increasing
separation times the fringes are observed to first wash out and
subsequently revive.}\label{fig:expramsfringe}
\end{center}
\end{figure}

As described in section \ref{sec:theodynamics} the dynamics of atoms
in our dipole trap is expected to affect the ability of an echo
pulse to balance the effect of two surrounding phase shifting light
pulses, as can indeed be observed experimentally in echo
spectroscopy. To illuminate the effect, we present in
Fig.~\ref{fig:expramsfringe} examples of Ramsey fringes as recorded
in the two-color probing scheme using the echo sequence of
Fig.~\ref{fig:sequence} for light pulse separations $\tau_1+\tau_2$
of 12, 250, and 500~$\rm \mu s$, respectively. The light pulses have
a duration of 2~$\mu s$ which is short compared to all other
relevant time scales in the experiment such as the trap oscillation
period of $\rm\sim1000~\mu s$. For the traces shown, $t_1$ was kept
fixed at 1500~ms while $t_2$ was scanned between 500~ms and 2500~ms.
To enable the observation of Ramsey fringes in the time domain the
frequency of our microwave source was detuned by 3~kHz. The detuning
sets the undulation frequency of the Ramsey fringes which have their
maximum amplitude at $t_2=t_1=1500$~ms due to the echo rephasing
pulse.

The most intriguing aspect of Fig.~\ref{fig:expramsfringe} is the
collapse and revival of fringe visibility. For a time separation
$\rm\tau_1+\tau_2=12~\mu s$ between the light pulses which is short
compared to the trap frequency, the echo pulse is effective in
restoring the Ramsey fringe visibility: Since the light pulses are
so closely spaced that the atoms hardly have time to move in
between, the second pulse essentially undoes the inhomogeneous phase
imprint of the first. An upper bound on the amount of decoherence
can be estimated from reduction in fringe amplitude, which in this
case is 76\%. A completely different situation is encountered at
$\rm\tau_1+\tau_2=250~\mu s$ with a light pulse separation in the
vicinity of a quarter of a radial trap period. Here the atomic
ensemble develops a nonuniform phase as indicated in
Fig.~\ref{fig:2}b(i) and the Ramsey fringes for individual atoms
will in general not interfere constructively. Hence a certain
degradation in the ensemble fringe contrast happens. Finally, for a
pulse separation of $\rm\tau_1+\tau_2=500~\mu s$ in the vicinity of
half a radial trap period a Ramsey fringe revival happens. At the
time of the second light pulse an atom will find itself close to the
radial distance (from the trap center) it had when the first pulse
was applied and hence experience the same light intensity. Due to
the inverting effect of the interposed echo pulse a phase shift
cancelation will occur.

\subsubsection{Revival frequency versus trap power}
\begin{figure}[tb!]
\begin{center}
\includegraphics[width=\columnwidth]{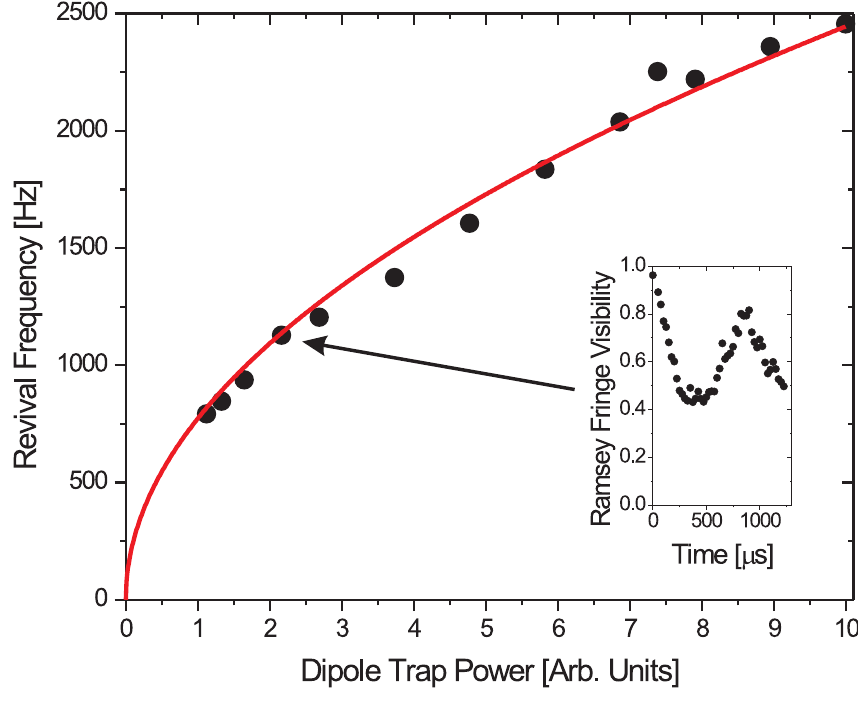}
\caption{Observed revival frequency of Ramsey fringes in echo
spectroscopy as a function of the dipole trapping power. A square
root scaling (line) describes the data well. Each data point was
extracted from a revival trace as shown in the inset as recorded
using the single color probing scheme.}\label{fig:expfreqdep}
\end{center}
\end{figure}

The revival of Ramsey fringe visibility, resulting from interference
between spatially imprinted phases in the atomic coherences, is
linked to the radial trap oscillation frequency. Measurements of
trap frequencies in cold atom experiments are typically performed by
exciting either monopole or dipole oscillation modes, or by driving
parametric losses when modulating the trap
\cite{Grimm2000,Boiron1998}. Using the fringe revival, we are able
to extract the trap frequency \textit{in situ} without actually
exciting motion
---  an atom is tagged in its \textit{internal} degrees of freedom according
to its position. Towards this end we simply measure the height of
the central Ramsey fringe in our echo sequence
Fig.~\ref{fig:sequence} as a function of the separation between the
two light shifting pulses. Hence, we keep $t_2=t_1$ fixed and vary
$\rm\tau_1+\tau_2$. As the height of the central Ramsey fringe
provides measure of the fringe visibility this quantity oscillates
in $\rm\tau_1+\tau_2$ at twice the trap frequency as discussed
above. Figure~\ref{fig:expfreqdep} shows the measured revival
frequency as function of the optical power of our dipole trapping
beam. The observed revival frequencies are described well by a
square root dependency on the optical power as expected from theory
\cite{Grimm2000}.

\subsubsection{Modeling the revivals}
\begin{figure}[b!]
\begin{center}
\includegraphics[width=\columnwidth]{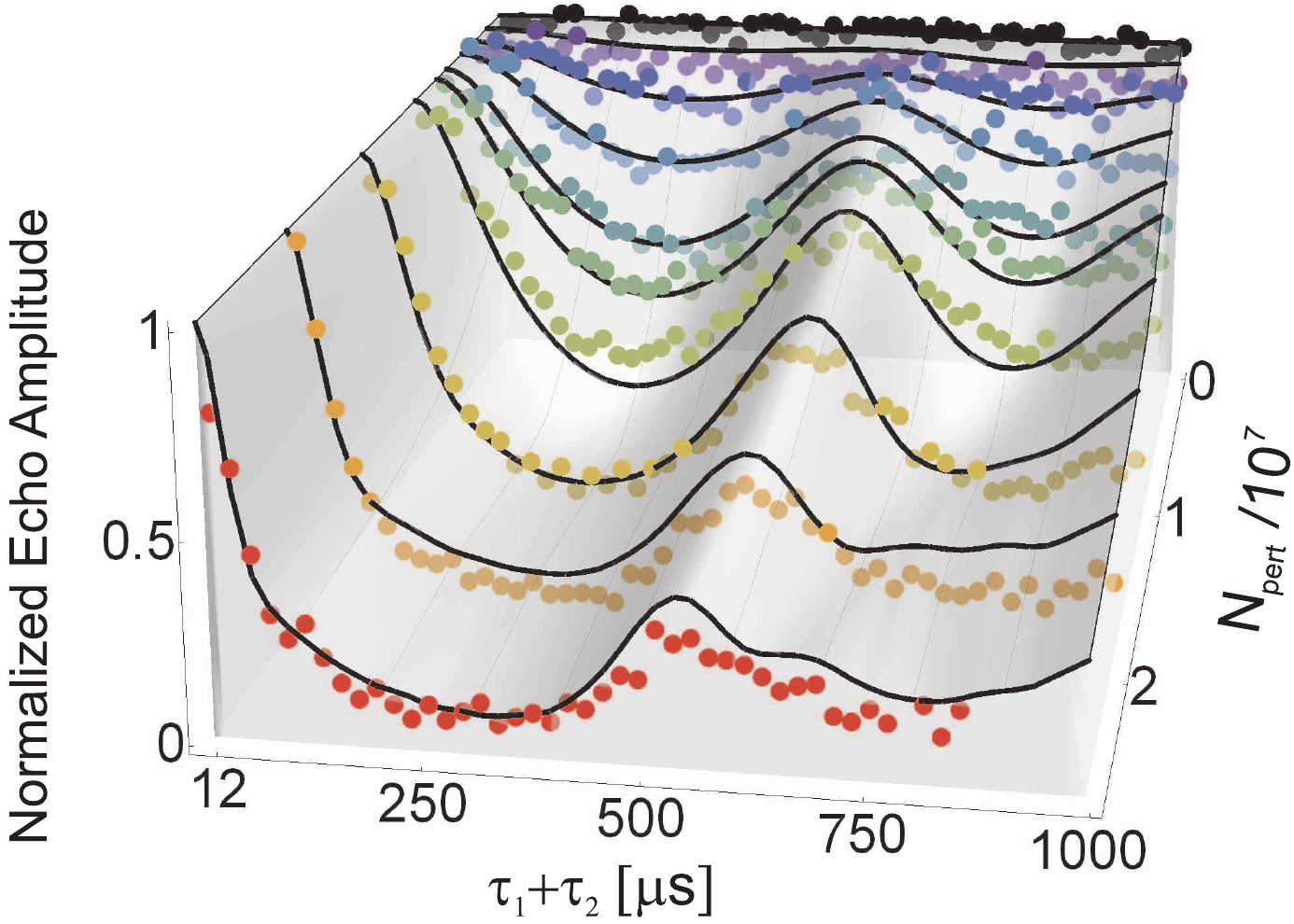}
\caption{Collapse and revival of the fringe amplitude in echo
spectroscopy. Filled circles shows the experimental normalized
fringe amplitude as function of the time separation $\tau_1+\tau_2$
between peturbing light pulses around the echo pulse for ten
different values of the number of perturbing photons $N_{\rm pert}$.
Full lines show the result of our numerical simulations of the
dynamics.}\label{fig:expdecaysrevivals}
\end{center}
\end{figure}
In a perfect harmonic trap the radial oscillation period is
independent of oscillation amplitude and complete revival of the
Ramsey fringe would hence be expected when $\rm\tau_1+\tau_2$ equals
half a trap period. This is clearly not the case for the revival
presented in the inset of Fig.~\ref{fig:expfreqdep}. A possible
explanation could be sought in the anharmonicity of the Gaussian
trapping potential. To investigate this in more detail we perform
numerical simulations of the phase evolution of $N=50,000$ particles
in a Gaussian trap when subjected to the echo spectroscopy sequence
shown in Fig.~\ref{fig:sequence} including a non-uniform light shift
from a Gaussian laser beam. During half a radial trapping period the
axial movement of even the most energetic, trapped particles will
only amount to a small fraction of the probe beam Rayleigh length
which characterizes the length scale for the light-atom interaction
volume. Furthermore, collisions between particles are expected to be
negligible for the low densities and short time scales involved
\cite{Mudrich2002}. In addition, we shall assume that the mean
kinetic energy for trapped particles is sufficiently large for the
radial frequency variation along the trap axis to be negligible
compared with the local frequency spread caused by the trap
potential anharmonicity. Hence, as an approximation, we shall
restrict our treatment to the 2D case of radial dynamics. The
details of the implementation of our numerical simulations is given
in appendix~\ref{app:spatdist}. An expected fringe amplitude
(relative to the unperturbed case) as function of time separation
between the two perturbing light pulses, $t\equiv \tau_1+\tau_2$, is
given by
\begin{equation}\label{eqfringecontrast}
\mathcal{F}(t)=\frac{\sum_{j=0}^Ne^{-\rho_j^2(t/2+t_2)/\gamma^2w_0^2}\cos{\phi_f^{(j)}(t)}}{\sum_{j=0}^Ne^{-\rho_j^2(t/2+t_2)/\gamma^2w_0^2}},
\end{equation}
where $\rho_j$ and $\phi_f^{(j)}$ are the radial position and echo
phase [cf. Eq.~(\ref{eqechophase})], respectively, of particle $j$
in the ensemble. Since the atomic ensemble is probed with a Gaussian
light beam, each particle contributes to the total signal by a
weight according to the intensity at its radial position at the time
of probing. Figure~\ref{fig:expdecaysrevivals} shows the fringe
amplitude versus light pulse separation time as recorded in echo
spectroscopy experiments for various perturbing light powers $P_{\rm
pert}$. We model our data using $\mathcal{F}(t)$ as provided by our
numerical simulations. The functional behavior of $\mathcal{F}(t)$
depends on the three parameters $T$, $\gamma$, and $\phi_0$, i.e.,
the cloud temperature prior to the separatrix truncation (see
appendix \ref{app:spatdist}), the probe to trap beam ratio, and the
peak phase shift. These parameters are adjusted to minimize the mean
square difference to our experimental data. Least squares are
obtained for $kT/U_0=1$, $\gamma=0.6$, and $\phi_0/N_{\rm
pert}=5.0\times10^{-7}$ which corresponds well to what would be
expected in the experiment. Fixing the free parameters of our
simulation code to these values, we plot the normalized echo
amplitude in Fig.~\ref{fig:expdecaysrevivals}. We note that our
simple 2D model gives an excellent over-all agreement between
experiment and simulation.

The model does not include decoherence with the result that all the
simulated curves in Fig.~\ref{fig:expdecaysrevivals} indicate
perfect rephasing at $t=0$. In the experimental data it is evident
that rephasing is not perfect especially for the large probe powers.
However, the effect of trap dynamics on the fringe visibility
clearly dominates over the decoherence for most pulse separations
and it is unclear how much the, in principle, reversible dephasing
contributes to the fringe reduction at the smallest pulse separation
of 12$\mu s$. Thus, measurements provide an upper bound on the
(irreversible) decoherence

\section{Discussion}
In summary, we have reported on the manifestation of motional
dynamics of trapped atoms in echo spectroscopy on the Cs clock
transition. The movement of particles implies that the ac Stark
shifts from two identical light pulses generally do not cancel
completely on the application of an in-between Hahn echo. Rather,
the echo fringe amplitude is observed to decrease with increasing
separation of the two light pulses. For obtaining
squeezing on the clock transition via dispersive measurements this
is unfortunate for two reasons. First, the effect limits the
applicability of echo techniques to restore the adverse
inhomogeneous light shift effects of a strong off-resonant probe
pulse. Second, using echo spectroscopy to gauge the atomic
decoherence resulting from spontaneously scattered probe photons
when probing an ensemble is not straight forward. Obviously, an
understanding of the effect of motional dynamics is important when
dispersive measurements are used for quantum state engineering.

An alternative approach to eliminate inhomogeneous differential
light shift of the clock states induced by a probe pulse is to apply
a simultaneous light pulse from a second laser \cite{Kaplan2004}.
The frequency of the second probe laser should be chosen so that the
differential ac Stark shifts introduced by each probe laser exactly
cancel. In the present configuration, where the two probe beams
enter the Mach Zehnder interferometer via the same input port, the
detunings are pegged such that simultaneous application of the two
probes on a coherent superposition state yields a zero mean phase
shift. At these detunings the ac Stark shift imposed on each clock
state add in sign and instead of cancelling the differential light
shift, the two-color probing scheme essentially doubles it. However,
injecting the two beams via \textit{different} ports of the
interferometer input beam splitter establishes a scheme where the
detunings can be chosen to eliminate the differential light shift
and achieve a state sensitive, balanced interferometer signal at the
same time \cite{saffmaninprep}. We are presently in the course of
reconfiguring our experimental setup to this favourable scheme. Echo
spectroscopy experiments along the lines the present paper should
then easily indicate if light shift cancelation has been
accomplished.

We recall that the primary motivation for our experimental efforts
is to demonstrate squeezing on the clock transition. Here a balanced
interferometer measurement of an ensemble in an equal clock state
superposition will be used to predict the outcome of a subsequent
measurement beyond the standard quantum limit
\cite{Windpassinger2008,Kuzmich1998,Oblak2005}. The present paper
and ref. \cite{pwarxiv1} have investigated some adverse effects on
the collective ensemble state which may accompany these
``nondestructive" measurements and discussed routes to minimize
them. Even when applying these strategies to eliminate inhomogeneous
light shifts, the atomic trap dynamics will lead to retardation
effects when comparing two consecutive measurements on an ensemble.
Particle motion and a Gaussian probe beam profile implies that an
atom in general contributes to the interferometer signal by
different weights in the two measurements. A fuller treatment of
this is beyond the scope of this paper, but it is expected that the
degree of squeezing (i.e. the amount of correlation between the two
pulses) is going the diminish and increase analogous to the collapse
and revival of echo fringe amplitude as reported on here.

\begin{acknowledgement}
We thank J\"{o}rg Helge M\"{u}ller for stimulating discussions. This work
was funded by the Danish National Research Foundation, as well as
the EU grants QAP, COVAQUIAL and CAUAC. N.K. acknowledges the
support of the Danish National Research Council through a Steno
Fellowship.
\end{acknowledgement}
\begin{appendix}
\section{Methods for numerical simulations}
\label{app:spatdist} 
\begin{figure}[tb!]
\begin{center}
\includegraphics[width=0.7\columnwidth]{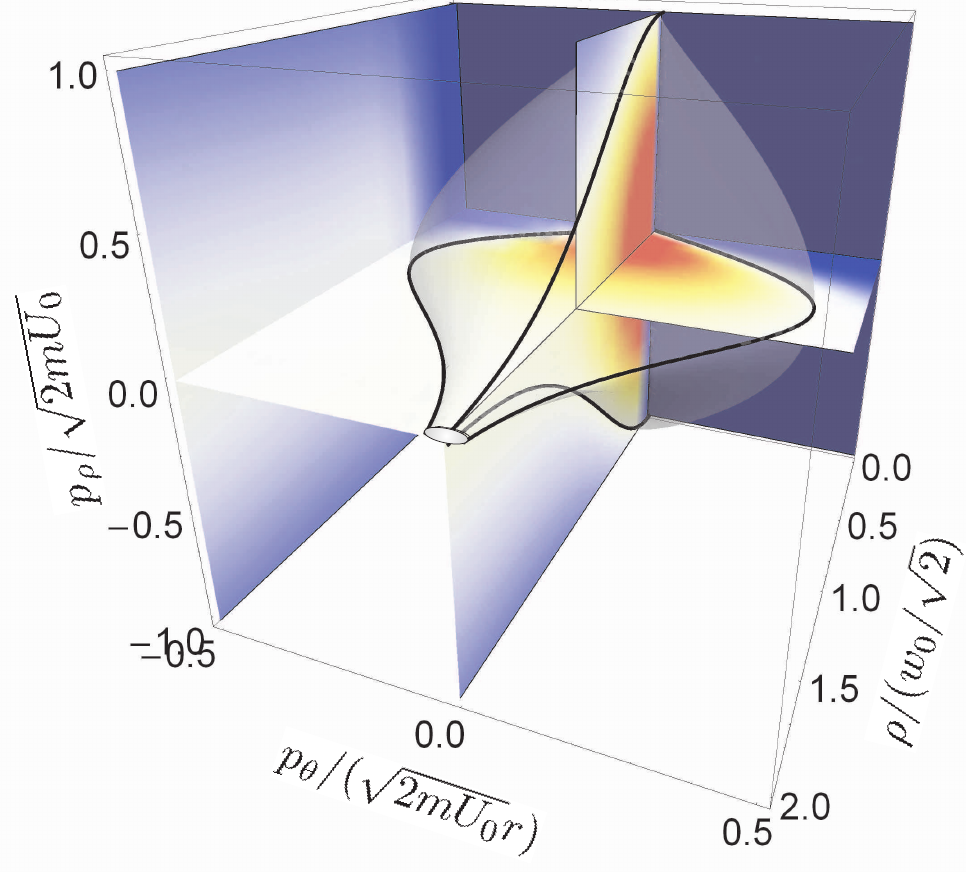}
\caption{Illustration of the Gaussian potential separatrix of bound
motion projected onto $(\rho,p_\rho,p_\theta)$ phase-space.
Particles inside the separatrix are trapped and form the starting
point for our numerical simulations.
 }\label{fig:separatrix}
\end{center}
\end{figure}

We obtain the initial conditions for our
numerical simulations by considering a canonical ensemble of
particles with mass $m$ at a temperature $T$. Using polar
coordinates $(\rho,\theta)$, the phase space density is
\begin{equation}
\sigma({\rho,\theta,p_{\rho}},p_{\theta})\propto
e^{-[({p_{\rho}^2+p_{\theta}^2 /\rho^2})/2m+V({\rho})]/kT},
\end{equation}
 where $k$ is the Boltzmann constant and $p_{\rho}$ and $p_{\rho}$ are the conjugate momenta to $\rho$ and $\theta$, respectively \cite{Reif1985}. For
the Gaussian trap the confinement potential is $V({\rho})=-
U_0e^{-(2\rho^2/w_0^2)}$, which has the (finite) trap depth $U_0$
and depends only on the radial distance $\rho$ to the trap center at
the origin. We want to restrict ourselves to particles inside the
separatrix of stable motion, i.e., we disregard unbound particles
for which $({p_{\rho}^2+p_{\theta}^2 /\rho^2})/2m+V({\rho})>0$ (see
Fig.~\ref{fig:separatrix}). Integrating over particles inside the
separatrix yields a spatial density distribution
\begin{equation}
n({\bf r})\propto\left[e^{-V(\rho)/kT}-1\right]\propto\left\{
\begin{array}{lcc}\exp\left(-\frac{U_0}{kT}\frac{2\rho^2}{w_0^2}\right)&,&U_0\gg kT \\ \exp\left(-\frac{2\rho^2}{w_0^2}\right)&,&U_0 \ll kT \\
\end{array}\right.,
\end{equation}
which for all practical purposes can be assumed to vanish for
$\rho\gtrsim w_0$. Applying a finite cut-off radius $\rho_0\gtrsim
w_0$, we assign random radii to the particles of our ensemble
according to a probability density $\propto \rho e^{-V(\rho)/kT}$
using the rejection method \cite{Press1986} on the interval
$[0...\rho_0]$. Next we assign the canonical momenta $p_\rho$ and
$p_\theta$ for each particle in correspondence with normal
distributions of widths $\sqrt{kT}$ and $\rho\sqrt{kT}$,
respectively. We discard untrapped particles as described above and
tag each trapped particle $j$ with a phase
$\phi_i^{(j)}=\phi_0\exp(-2\rho_j^2/\gamma^2w_0^2)$, where $\gamma$
is the ratio between the waists of the trapping laser and the
perturbing laser and $\phi_0$ is the peak phase shift. We finally
integrate the equations of motion using the Verlet method to obtain
$\rho_j(t)$ from which the echo phase
\begin{equation}\label{eqechophase}
\phi_f^{(j)}(t)=\phi_i^{(j)}-\phi_0\exp(-2\rho_j^2(t)/\gamma^2w_0^2)
\end{equation}
can be calculated.
\end{appendix}

 \bibliographystyle{epj}

\begin{thebibliography}{21}

\bibitem{Chaudhury2006}
S.~Chaudhury, G.A. Smith, K.~Schulz, P.S. Jessen, Phys. Rev. Lett.
  \textbf{96}(4), 043001 (2006)

\bibitem{Windpassinger2008}
P.J. Windpassinger, D.~Oblak, P.G. Petrov, M.~Kubasik, M.~Saffman,
C.L.G.
  Alzar, J.~Appel, J.H. M\"{u}ller, N.~Kj{\ae}rgaard, E.S. Polzik, Phys. Rev.
  Lett. \textbf{{100}}, 103601 ({2008})

\bibitem{Kuzmich1998}
A.~Kuzmich, N.P. Bigelow, L.~Mandel, Europhys. Lett. \textbf{42},
481 (1998)

\bibitem{Oblak2005}
D.~Oblak, P.G. Petrov, C.L. Garrido~Alzar, W.~Tittel, A.K.
Vershovski, J.K.
  Mikkelsen, J.L. S{\o}rensen, E.S. Polzik, Phys. Rev. A \textbf{71}, 043807
  (2005)

\bibitem{pwarxiv1}
P.J. Windpassinger, D.~Oblak, U.B. Hoff, J.~Appel, N.~Kj{\ae}rgaard,
E.S.
  Polzik, New. J. Phys. \textbf{10}, 053032 (2008)

\bibitem{Vanier1989}
J.~Vanier, C.~Audoin, \emph{The Quantum Physics of Atomic Frequency
Standards}
  (Adam Hilger, Bristol, 1989)

\bibitem{Featonby1998}
P.D. Featonby, C.L. Webb, G.S. Summy, C.J. Foot, K.~Burnett, J.
Phys. B
  \textbf{31}, 375 (1998)

\bibitem{HAHN1950}
E.L. Hahn, Phys. Rev. \textbf{80}, 580 (1950)

\bibitem{Kuhr2005}
S.~Kuhr, W.~Alt, D.~Schrader, I.~Dotsenko, Y.~Miroshnychenko,
  A.~Rauschenbeutel, D.~Meschede, Phys. Rev. A \textbf{72}, 023406 (2005)

\bibitem{Andersen2003}
M.F. Andersen, A.~Kaplan, N.~Davidson, Phys. Rev. Lett. \textbf{90},
023001
  (2003)

\bibitem{Ozeri2005}
R.~Ozeri, C.~Langer, J.D. Jost, B.~DeMarco, A.~Ben-Kish, B.R.
Blakestad,
  J.~Britton, J.~Chiaverini, W.M. Itano, D.B. Hume et~al., Phys. Rev. Lett.
  \textbf{95}, 030403 (2005)

\bibitem{Lengwenus2007}
A.~Lengwenus, J.~Kruse, M.~Volk, W.~Ertmer, G.~Birkl, Appl. Phys. B
  \textbf{86}, 377 (2007)

\bibitem{Petrov2007a}
P.G. Petrov, D.~Oblak, C.L. Garrido~Alzar, N.~Kj{\ae}rgaard, E.S.
Polzik, Phys.
  Rev. A \textbf{75}, 033803 (2007)

\bibitem{Hammerer2004}
K.~Hammerer, K.~M{\o}lmer, E.S. Polzik, J.I. Cirac, Phys. Rev. A
  \textbf{70}(4), 044304 (2004)

\bibitem{Grimm2000}
R.~Grimm, M.~Weidemuller, Y.B. Ovchinnikov, Adv. At. Mol. Opt. Phys.
  \textbf{42}, 95 (2000)

\bibitem{Boiron1998}
D.~Boiron, A.~Michaud, J.M. Fournier, L.~Simard, M.~Sprenger,
G.~Grynberg,
  S.~C., Phys. Rev. A \textbf{57}, R4106 (1998)

\bibitem{Mudrich2002}
M.~Mudrich, S.~Kraft, K.~Singer, R.~Grimm, A.~Mosk,
M.~Weidem\"{u}ller, Phys.
  Rev. Lett. \textbf{88}, 253001 (2002)

\bibitem{Kaplan2004}
A.~Kaplan, M.F. Andersen, N.~Davidson, Phys. Rev. A \textbf{66},
045401 (2004)

\bibitem{saffmaninprep}
M.~Saffman~{\it et al.}, in prep.  (2008)

\bibitem{Reif1985}
F.~Reif, \emph{Fundamentals of Statistical and Thermal Physics}
(McGraw-Hill,
  Singapore, 1985)

\bibitem{Press1986}
W.H. Press, B.P. Flannery, S.A. Teukolsky, W.T. Vetterling,
\emph{Numerical
  Recipes} (Cambridge University Press, Cambridge, 1986)

\end{thebibliography}

\end{document}